\documentstyle[11pt,epsf]{article}
\begin{document}
\newcommand {\ee}{\end{equation}}
\newcommand {\bea}{\begin{eqnarray}}
\newcommand {\eea}{\end{eqnarray}}
\newcommand {\nn}{\nonumber \\}
\newcommand {\Tr}{{\rm Tr\,}}
\newcommand {\tr}{{\rm tr\,}}
\newcommand {\e}{{\rm e}}
\newcommand {\etal}{{\it et al.}}
\newcommand {\m}{\mu}
\newcommand {\n}{\nu}
\newcommand {\pl}{\partial}
\newcommand {\p} {\phi}
\newcommand {\vp}{\varphi}
\newcommand {\vpc}{\varphi_c}
\newcommand {\al}{\alpha}
\newcommand {\be}{\beta}
\newcommand {\ga}{\gamma}
\newcommand {\Ga}{\Gamma}
\newcommand {\x}{\xi}
\newcommand {\ka}{\kappa}
\newcommand {\la}{\lambda}
\newcommand {\La}{\Lambda}
\newcommand {\si}{\sigma}
\newcommand {\Si}{\Sigma}
\newcommand {\th}{\theta}
\newcommand {\Th}{\Theta}
\newcommand {\om}{\omega}
\newcommand {\Om}{\Omega}
\newcommand {\ep}{\epsilon}
\newcommand {\vep}{\varepsilon}
\newcommand {\na}{\nabla}
\newcommand {\del}  {\delta}
\newcommand {\Del}  {\Delta}
\newcommand {\mn}{{\mu\nu}}
\newcommand {\ls}   {{\lambda\sigma}}
\newcommand {\ab}   {{\alpha\beta}}
\newcommand {\half}{ {\frac{1}{2}} }
\newcommand {\third}{ {\frac{1}{3}} }
\newcommand {\fourth} {\frac{1}{4} }
\newcommand {\sixth} {\frac{1}{6} }
\newcommand {\sqg} {\sqrt{g}}
\newcommand {\fg}  {\sqrt[4]{g}}
\newcommand {\invfg}  {\frac{1}{\sqrt[4]{g}}}
\newcommand {\sqZ} {\sqrt{Z}}
\newcommand {\gbar}{\bar{g}}
\newcommand {\sqk} {\sqrt{\kappa}}
\newcommand {\sqt} {\sqrt{t}}
\newcommand {\reg} {\frac{1}{\epsilon}}
\newcommand {\fpisq} {(4\pi)^2}
\newcommand {\Lcal}{{\cal L}}
\newcommand {\Ocal}{{\cal O}}
\newcommand {\Dcal}{{\cal D}}
\newcommand {\Ncal}{{\cal N}}
\newcommand {\Mcal}{{\cal M}}
\newcommand {\scal}{{\cal s}}
\newcommand {\Dvec}{{\hat D}}   
\newcommand {\dvec}{{\vec d}}
\newcommand {\Evec}{{\vec E}}
\newcommand {\Hvec}{{\vec H}}
\newcommand {\Vvec}{{\vec V}}
\newcommand {\Btil}{{\tilde B}}
\newcommand {\ctil}{{\tilde c}}
\newcommand {\Ftil}{{\tilde F}}
\newcommand {\Stil}{{\tilde S}}
\newcommand {\Ztil}{{\tilde Z}}
\newcommand {\altil}{{\tilde \alpha}}
\newcommand {\betil}{{\tilde \beta}}
\newcommand {\latil}{{\tilde \lambda}}
\newcommand {\ptil}{{\tilde \phi}}
\newcommand {\Ptil}{{\tilde P}}
\newcommand {\natil} {{\tilde \nabla}}
\newcommand {\ttil} {{\tilde t}}
\newcommand {\Rhat}{{\hat R}}
\newcommand {\Shat}{{\hat S}}
\newcommand {\shat}{{\hat s}}
\newcommand {\Dhat}{{\hat D}}   
\newcommand {\Vhat}{{\hat V}}   
\newcommand {\xhat}{{\hat x}}
\newcommand {\Zhat}{{\hat Z}}
\newcommand {\Gahat}{{\hat \Gamma}}
\newcommand {\nah} {{\hat \nabla}}
\newcommand {\gh}  {{\hat g}}
\newcommand {\labar}{{\bar \lambda}}
\newcommand {\cbar}{{\bar c}}
\newcommand {\bbar}{{\bar b}}
\newcommand {\Bbar}{{\bar B}}
\newcommand {\psibar}{{\bar \psi}}
\newcommand {\chibar}{{\bar \chi}}
\newcommand {\bbartil}{{\tilde {\bar b}}}
\newcommand  {\vz}{{v_0}}
\newcommand  {\ez}{{e_0}}
\newcommand  {\rhc}{{\rho_c}}
\newcommand {\intfx} {{\int d^4x}}
\newcommand {\inttx} {{\int d^2x}}
\newcommand {\change} {\leftrightarrow}
\newcommand {\ra} {\rightarrow}
\newcommand {\larrow} {\leftarrow}
\newcommand {\ul}   {\underline}
\newcommand {\pr}   {{\quad .}}
\newcommand {\com}  {{\quad ,}}
\newcommand {\q}    {\quad}
\newcommand {\qq}   {\quad\quad}
\newcommand {\qqq}   {\quad\quad\quad}
\newcommand {\qqqq}   {\quad\quad\quad\quad}
\newcommand {\qqqqq}   {\quad\quad\quad\quad\quad}
\newcommand {\qqqqqq}   {\quad\quad\quad\quad\quad\quad}
\newcommand {\qqqqqqq}   {\quad\quad\quad\quad\quad\quad\quad}
\newcommand {\lb}    {\linebreak}
\newcommand {\nl}    {\newline}

\newcommand {\vs}[1]  { \vspace*{#1 cm} }

\newcommand {\MPL}  {Mod.Phys.Lett.}
\newcommand {\NP}   {Nucl.Phys.}
\newcommand {\PL}   {Phys.Lett.}
\newcommand {\PR}   {Phys.Rev.}
\newcommand {\PRL}   {Phys.Rev.Lett.}
\newcommand {\CMP}  {Commun.Math.Phys.}
\newcommand {\JMP}  {Jour.Math.Phys.}
\newcommand {\AP}   {Ann.of Phys.}
\newcommand {\PTP}  {Prog.Theor.Phys.}
\newcommand {\NC}   {Nuovo Cim.}
\newcommand {\CQG}  {Class.Quantum.Grav.}


\font\smallr=cmr5
\def\ocirc#1{#1^{^{{\hbox{\smallr\llap{o}}}}}}
\def\ogamma{\ocirc{\gamma}{}}
\def\oM{{\buildrel {\hbox{\smallr{o}}} \over M}}
\def\osigma{\ocirc{\sigma}{}}

\def\overleftrightarrow#1{\vbox{\ialign{##\crcr
 $\leftrightarrow$\crcr\noalign{\kern-1pt\nointerlineskip}
 $\hfil\displaystyle{#1}\hfil$\crcr}}}
\def\overnab{{\overleftrightarrow\nabslash}}

\def\va{{a}}
\def\vb{{b}}
\def\vc{{c}}
\def\tilpsi{{\tilde\psi}}
\def\tbpsi{{\tilde{\bar\psi}}}

\def\Dslash{{}\hbox{\hskip2pt\vtop
 {\baselineskip23pt\hbox{}\vskip-24pt\hbox{/}}
 \hskip-11.5pt $D$}}
\def\nabslash{{}\hbox{\hskip2pt\vtop
 {\baselineskip23pt\hbox{}\vskip-24pt\hbox{/}}
 \hskip-11.5pt $\nabla$}}
\def\xislash{{}\hbox{\hskip2pt\vtop
 {\baselineskip23pt\hbox{}\vskip-24pt\hbox{/}}
 \hskip-11.5pt $\xi$}}
\def\leftnabla{{\overleftarrow\nabla}}

\def\delL{{\delta_{LL}}}
\def\delG{{\delta_{G}}}
\def\delc{{\delta_{cov}}}

\newcommand {\sqxx}  {\sqrt {x^2+1}}   
\newcommand {\gago}  {\gamma_5}
\newcommand {\Ktil}  {{\tilde K}}
\newcommand {\Ltil}  {{\tilde L}}
\newcommand {\Qtil}  {{\tilde Q}}
\newcommand {\Rtil}  {{\tilde R}}
\newcommand {\Kbar}  {{\bar K}}
\newcommand {\Lbar}  {{\bar L}}
\newcommand {\Qbar}  {{\bar Q}}
\newcommand {\Pp}  {P_+}
\newcommand {\Pm}  {P_-}
\newcommand {\GfMp}  {G^{5M}_+}
\newcommand {\GfMpm}  {G^{5M'}_-}
\newcommand {\GfMm}  {G^{5M}_-}
\newcommand {\Omp}  {\Omega_+}    
\newcommand {\Omm}  {\Omega_-}
\def\Aslash{{}\hbox{\hskip2pt\vtop
 {\baselineskip23pt\hbox{}\vskip-24pt\hbox{/}}
 \hskip-11.5pt $A$}}
\def\Rslash{{}\hbox{\hskip2pt\vtop
 {\baselineskip23pt\hbox{}\vskip-24pt\hbox{/}}
 \hskip-11.5pt $R$}}
\def\kslash{
{}\hbox       {\hskip2pt\vtop
                   {\baselineskip23pt\hbox{}\vskip-24pt\hbox{/}}
               \hskip-8.5pt $k$}
           }    
\def\qslash{
{}\hbox       {\hskip2pt\vtop
                   {\baselineskip23pt\hbox{}\vskip-24pt\hbox{/}}
               \hskip-8.5pt $q$}
           }    
\def\dslash{
{}\hbox       {\hskip2pt\vtop
                   {\baselineskip23pt\hbox{}\vskip-24pt\hbox{/}}
               \hskip-8.5pt $\partial$}
           }    
\def\dbslash{{}\hbox{\hskip2pt\vtop
 {\baselineskip23pt\hbox{}\vskip-24pt\hbox{$\backslash$}}
 \hskip-11.5pt $\partial$}}
\def\Kbslash{{}\hbox{\hskip2pt\vtop
 {\baselineskip23pt\hbox{}\vskip-24pt\hbox{$\backslash$}}
 \hskip-11.5pt $K$}}
\def\Ktilbslash{{}\hbox{\hskip2pt\vtop
 {\baselineskip23pt\hbox{}\vskip-24pt\hbox{$\backslash$}}
 \hskip-11.5pt ${\tilde K}$}}
\def\Ltilbslash{{}\hbox{\hskip2pt\vtop
 {\baselineskip23pt\hbox{}\vskip-24pt\hbox{$\backslash$}}
 \hskip-11.5pt ${\tilde L}$}}
\def\Qtilbslash{{}\hbox{\hskip2pt\vtop
 {\baselineskip23pt\hbox{}\vskip-24pt\hbox{$\backslash$}}
 \hskip-11.5pt ${\tilde Q}$}}
\def\Rtilbslash{{}\hbox{\hskip2pt\vtop
 {\baselineskip23pt\hbox{}\vskip-24pt\hbox{$\backslash$}}
 \hskip-11.5pt ${\tilde R}$}}
\def\Kbarbslash{{}\hbox{\hskip2pt\vtop
 {\baselineskip23pt\hbox{}\vskip-24pt\hbox{$\backslash$}}
 \hskip-11.5pt ${\bar K}$}}
\def\Lbarbslash{{}\hbox{\hskip2pt\vtop
 {\baselineskip23pt\hbox{}\vskip-24pt\hbox{$\backslash$}}
 \hskip-11.5pt ${\bar L}$}}
\def\Rbarbslash{{}\hbox{\hskip2pt\vtop
 {\baselineskip23pt\hbox{}\vskip-24pt\hbox{$\backslash$}}
 \hskip-11.5pt ${\bar R}$}}
\def\Qbarbslash{{}\hbox{\hskip2pt\vtop
 {\baselineskip23pt\hbox{}\vskip-24pt\hbox{$\backslash$}}
 \hskip-11.5pt ${\bar Q}$}}
\def\Acalbslash{{}\hbox{\hskip2pt\vtop
 {\baselineskip23pt\hbox{}\vskip-24pt\hbox{$\backslash$}}
 \hskip-11.5pt ${\cal A}$}}

\begin{flushright}
April 2001
\\
hep-th/0103211\\
YITP-SB-01-11
\end{flushright}

\vspace{0.5cm}

\begin{center}
{\Large\bf 
Some Properties of\\
Pole Solution in Six Dimensions}

\vspace{1.5cm}
{\large Shoichi ICHINOSE
          \footnote{
On leave of absence from 
Lab. of Physics, SFNS, University of Shizuoka,
Yada 52-1, Shizuoka 422-8526, Japan (Address after April 1, 2001).\nl          
E-mail address:\ ichinose@u-shizuoka-ken.ac.jp
                  }
}
\vspace{1cm}

{\large C.N. Yang Institute for Theoretical Physics\\
State University of New York at Stony Brook \\
Stony Brook, NY 11794-3840, USA}


\end{center}
\vfill
{\large Abstract}\nl
A solution with the pole configuration in six dimensions is
analyzed both analytically and numerically.
It is a dimensional reduction model of Radall-Sundrum type.
The soliton configuration is induced by
the bulk Higgs mechanism. 
The boundary condition is systematically solved up to
the 6th order.
The Riemann curvature is finite everywhere. 
An exact solution for the no potential case
is also presented.

\vspace{0.5cm}
PACS NO:\ 
11.27.+d 
04.50.+h 
11.10.Kk 
11.25.Mj 
12.10.-g 
04.20.Ex 
04.25.-g 
\nl
Key Words:\ Boundary condition, Randall-Sundrum model, 
Dimensional reduction, Brane World, Six dimensions, Pole solution
\newpage
\section{Introduction}
Seeking a realistic model of our world, from the 
higher-dimensional standpoint, 
the brane-world approach is now vigorously
envestigated in various ways.
It started with the wall configuration in 5 dimensions(D)
space-time\cite{RS9905,RS9906}. Basically the brane-world physics
makes use of the soliton configuration. 
We can explore new physics, including the beyond-standard model, 
by making use of the distinguished properties of soliton,
such as the localized zero modes, stable configuration,
singularity-free behavior due to the extendedness, etc..
From the reason of improving 
some difficulties in 5D model\cite{KL0001,GL0003,MN0007}, 
or of getting more general
aspects of the brane-world physics, 
further-higher-dimensional models ( string-like object
in 6D, monopole configuration in 7D, $\cdots$)
are also envestigated\cite{Oda0006,DS0008,DS0008b,GS0004,OV0003}.
In this paper, we examine a 6D model.

In the original model, 3-brane(s) is introduced, by hand, as a
$\del$-function distribution in the extra space. We do not
take such approach in order to seek a non-singular solution.
We induce the same configuration as a soliton (kink) solution.
This situation is very similar to the relation between the Dirac
 and 'tHooft-Polyakov monopoles. The latter is the soliton
(in the gauge+Higgs system) interpretation of the former
system where the gauge potential has $\del$-function distribution
in the plane perpendicular to the Dirac string. We seek a soliton
solution in the 6D gravity + Higgs system. 
The main purpose is to establish the pole solution
\cite{SI0012}. The most crucial point is 
to confirm the convergence
of some infinite series appearing in the solution. It guarantees
the boundary condition. In \cite{SI0012}
they are solved up to the 2nd order approximation. Here we show
them up to
the 6th order. The result strongly shows that 
the series converge and the boundary
conditions are sufficiently satisfied.

\section{The Dimensional Reduction Model in Six Dimensions}

In \cite{SI0012} a solution with the pole configuration
in six dimensions is presented. The model is the 6D gravity
coupled with the Higgs fields.
\begin{eqnarray}
S[G_{AB},\Phi]=\int d^6X\sqrt{-G} (-\half M^4\Rhat
- G^{AB}\pl_A\Phi^*\pl_B\Phi-V(\Phi^*,\Phi))\com\nn
V(\Phi^*,\Phi)=\frac{\la}{4}(|\Phi|^2-{v_0}^2)^2+\La\com
\label{model1}
\end{eqnarray}
where $(X^A)\equiv (x^\m,\rho,\vp), \m=0,1,2,3$. 
$x^\m$'s are regarded as our world coordinates, whereas  
$(\rho,\vp)$ the extra ones(\ $0\leq \rho <\infty,\ 0\leq \vp <2\pi$).
$G_{AB}$ is the 6D metric field, 
$\Phi^*$ and $\Phi$ are the complex scalars (Higgs fields). 
$M(>0)$ is the 6D Planck mass.  
$\la(>0),\vz(>0)$ and $\La$ in the potential $V$ are
{\it vacuum parameters}. 
Expecting Randall-Sundrum type dimensional reduction, we assume
the following form as the line element.
\begin{eqnarray}
{ds}^2=\e^{-2\si(\rho)}\eta_\mn dx^\m dx^\n+{d\rho}^2
+\rho^2\e^{-2\om(\rho)}d\vp^2\com\nn
0\leq \rho<\infty\com\q 0\leq\vp<2\pi\com
\label{model2}
\end{eqnarray}
where $\eta_\mn=\mbox{diag}(-1,1,1,1)$. 
This is a natural 6D 
minimal-extension of the original 5D model by Randall-Sundrum
\cite{RS9905}.  Two "warp" factors, $\e^{-2\si(\rho)}$ and
$\e^{-2\om(\rho)}$, appear. 
For the fixed $\rho$ case ($d\rho=0$), the metric
is the Weyl transformation of the product-space, 
the 4D Minkowski $\times$\  the circle $S^1$.
The coordinate $\rho$ can be regarded as the {\it scaling} parameter.
We require the periodicity in $\vp$ for $\Phi(\rho,\vp)$ 
and assume the form: 
$\Phi= P(\rho)\e^{im\vp}\ (m=0,\pm1,\pm2,\cdots)$.
The 6D classical field equations of (\ref{model1}) are obtained as
\begin{eqnarray}
 3\si''-\frac{\si'}{\rho}+\si'\om'+\om''-(\om')^2
+2\frac{\om'}{\rho}
=2M^{-4}{P'}^2\com\nn
-16{\si'}^2+4\frac{\si'}{\rho}-4\si'\om'+4\si''
=2M^{-4}V\com\nn
4\si''-10{\si'}^2
=M^{-4}({P'}^2-m^2\frac{\e^{2\om}}{\rho^2}P^2+V)\pr\label{sol5}
\end{eqnarray}
We take a unit $M=1$ in the following for simplicity.
\footnote{
This is allowed by 
the following invariance of (\ref{model1}) or (\ref{sol5})
, due to the simple dimensional counting:\  $
\Phi\ra M^2\Phi,\ (P\ra M^2P)\ ,\ 
\vz\ra M^2\vz,\ \la\ra M^{-2}\la\ , 
\La\ra M^6\La\ ,\ x^\m\ra M^{-1}x^\m\ ,\ 
\rho\ra M^{-1}\rho\ .
$
}

\section{An Exact Solution for the No Potential Case}
In order to see the structure of the solution, it is useful
to consider a simple case, that is, the no potential case:\ 
$\la=\La=0\ (V(\Phi^*,\Phi)=0)$. Even in this case, 
the scalars are still coupled 
with gravity. In this section we stress those properties which are
shared with the general case of Sec.4. 
We consider the $m=0$ ({\it no flux}) solution. 
The following exact solution of the rational
function type can be obtained.
\begin{eqnarray}
\si'(\rho)=-\frac{1}{A\rho+B},\ 
\om'(\rho)=\frac{4\rho+B}{\rho (A\rho+B)},\ 
{P(\rho)'}^2=\frac{2(2A-5)}{(A\rho+B)^2}\com
\label{exact1}
\end{eqnarray}
where $A(\geq 5/2)$ and $B$ are {\it integration constants}. 
$A$ is a dimensionless constant, 
whereas $B$ is a dimensional one ( the dimension of length ).
The lower bound for $A$ comes from the {\it positivity}
of ${P'}^2$. 
This is a 2-parameters family solution (for $\si',\om'$, and $P'$)
at this stage.
Some special cases are
i)\ $A=5/2,\ P'=0$;\ 
ii)\ $B=0,\ -4\si'=\om'=(4/A)\times \rho^{-1}, 
{P'}^2=(2(2A-5)/A^2)\times \rho^{-2}$;\ 
iii)\ $A=4,\ \om'=\rho^{-1}$;\ 
iv)\ $A\ra\infty$, $\si'=0$, $\om'=0$,$P'=0$. 
The integration constants are fixed by a required 
{\it boundary condition}. 

Due to the condition $A\geq 5/2$, we do {\it not} find the solution
which satisfies the RS-type configuration:\ $\si'\ra$ const.($>0$) 
as $\rho\ra \infty$. However, if we restrict the region of $\rho$
as $|B|\gg A\rho$, then we have 
$\si'\sim -1/B,\ \om'\sim (1/\rho)-(A-4)/B,\ {P'}^2\sim 2(2A-5)/B^2$.
This looks a RS-type metric although we {\it cannot} 
take $\rho\ra\infty$. 

The final solution for $\si,\ \om$ and $P$ are given as
\begin{eqnarray}
\si(\rho)=-\frac{1}{A}\ln\, \frac{|A\rho+B|}{D},\ 
\om(\rho)=\ln \frac{|\rho|\cdot |A\rho+B|^{-\frac{A-4}{A}}}
{E^{\frac{4}{A}}},\nn 
P(\rho)=\pm\frac{\sqrt{2(2A-5)}}{A}\ln\, \frac{|A\rho+B|}{C}\com
\label{exact2}
\end{eqnarray}
where $C(>0),D(>0),E(>0)$ are another {\it integration constants}
with the dimension of length. 
They come from the symmetry of the {\it constant translation}
($\si\ra \si+\mbox{const.},\om\ra \om+\mbox{const.}, 
P\ra P+\mbox{const.}$) in eq.(\ref{sol5}) with $V=0$. 
This final
form (\ref{exact2}) is a 5-parameters family of solutions.

The line element is given by 
\begin{eqnarray}
{ds}^2=\left\{\frac{|A\rho+B|}{D}\right\}^{2/A}\eta_\mn dx^\m dx^\n+{d\rho}^2
+\left\{\frac{|A\rho+B|}{E}\right\}^{2(A-4)/A}\cdot E^2d\vp^2\ .\nn
0<\frac{2}{A}\leq \frac{4}{5}\com\q
-\frac{6}{5}\leq \frac{2(A-4)}{A}< 2\com
\label{exact3}
\end{eqnarray}
For the case of $A\ra\infty$ the space-time becomes
locally 6D Minkowski with the deficit angle
$\del=2\pi (1-A)=-\infty$. For the restricted region $|B|\gg A\rho$,
the above metric reduces to
\begin{eqnarray}
{ds}^2=\e^{\frac{2}{B}\rho}\left\{\frac{|B|}{D}\right\}^{2/A}
\eta_\mn dx^\m dx^\n+
{d\rho}^2
+\e^{\frac{2(A-4)}{B}\rho}\left\{\frac{|B|}{E}\right\}^{2(A-4)/A}
\cdot E^2d\vp^2\ .
\label{exact3b}
\end{eqnarray}
As a form, this is a RS-type metric although we {\it cannot} keep
it in $\rho\ra\infty$ region.

The 6D Riemann scalar curvature is obtained as
\begin{eqnarray}
\Rhat=-\frac{4(2A-5)}{(A\rho+B)^2}\leq 0\com
\label{exact4}
\end{eqnarray}
which is negative semi-definite. 
The absolute value goes to 0 as $\rho\ra\infty$. 
For $A=5/2$ (the case i) above), 
$\Rhat=0$ is valid everywhere. The metric in
this case
\begin{eqnarray}
A=\frac{5}{2}\com\nn
{ds}^2=\left\{\frac{|\frac{5}{2}\rho+B|}{D}\right\}^{4/5}
\eta_\mn dx^\m dx^\n
+{d\rho}^2
+\left\{\frac{E}{|\frac{5}{2}\rho+B|}\right\}^{6/5}\cdot E^2d\vp^2\ .
\label{exact5}
\end{eqnarray}
For $\infty>A>5/2$, we can see three cases:\ 
a) $B>0$, There is {\it no curvature singularity};\ 
b) $B=0$, The curvature is singular at $\rho=0$;\ 
c) $B<0$, The curvature singular at $\rho=-B/A$. 
As for the horizon, 
from the condition $0<\frac{2}{A}\leq \frac{4}{5}$, 
there is no horizon in the region $0\leq \rho <\infty$.
Hence, in b) and c), the singularities are naked ones.

The structure of the solution (\ref{exact3}) is similar
to the Kasner solution\cite{KSHM80} in 1+3 D. 
Ref.\cite{Ver99} explains the cosmic string solutions
in the Abelian Higgs model in 1+3 D. 
The two branches explained there, that is,
the cosmic string branch and the Melvin branch
look to have correspondence with the $A\ra\infty$ case and
$A=5/2$ case respectively.

\section{Pole Solution in Six Dimensions}
Now we consider for 
the case of the spontaneous breakdown (Higgs) potential:\  
$\la>0, \vz>0$, and $\La$ is general(at this stage).
The boundary
condition for $P$ should be taken as 
\begin{eqnarray}
\rho\ra\infty\com\q P(\rho)\ra +\vz
\com\label{sol6}
\end{eqnarray}
where we take the plus sign, using the freedom of
$\Phi \change -\Phi$ in (\ref{sol5}),
for the asymptotic value of $P$. 
We require (based on the analogy to 5D Randall-Sundrum
model\cite{RS9905,RS9906,SI00apr}) that 
$\si'$ and $ \om'$ go to  constants as $\rho\ra\infty$. 
Then we can deduce, using (\ref{sol5}), that
\begin{eqnarray}
m=0\mbox{  (no flux)}\com\nn
\rho\ra\infty\com\q
\si'\ra  \al \com\q \om'\ra \al\com\q
 \al=+\sqrt{\frac{-\La}{10}}M^{-2}
\pr\label{sol7}
\end{eqnarray}
This result says $\La$ should be taken negative: $\La<0$. 
We choose the plus sign as the asymptotic value
so that the Weyl scaling factors in (\ref{model2})
work as shrinking the proper distance as $\rho$ increases.
This choice guarantees the stableness of the system.
As for the behavior near $\rho=0$(ultra-violet region), 
we can take, in a consistent way
with (\ref{sol5}), as
\begin{eqnarray}
\rho\ra +0\com\q
\si'\ra s\rho^a\com\q  \om'\ra w\rho^b\com\q P\ra x\rho^c\nn
a=b=1,\ c=0\com\q s\neq 0,w\neq 0,x\neq 0\com\q\nn
s=\frac{\la}{16}(x^2-\vz^2)^2+\frac{\La}{4}\com
\label{sol8b}
\end{eqnarray}
with the additional possible case:\ 
$x=0$ with $c>1$.

We examine the following form of solutions\cite{SI0012}.
\begin{eqnarray}
\si'(\rho)=\al\sum_{n=0}^\infty\frac{c_{2n+1}}{(2n+1)!}\{\tanh (k\rho)\}^{2n+1}\com\nn
\om'(\rho)=\al\sum_{n=0}^\infty\frac{d_{2n+1}}{(2n+1)!}\{\tanh (k\rho)\}^{2n+1}\com\nn
P(\rho)=v_0\sum_{n=0}^\infty\frac{e_{2n}}{(2n)!}\{\tanh (k\rho)\}^{2n}\com
\label{sol9}
\end{eqnarray}
where   
$c$'s, $d$'s and $e$'s are coefficient-constants to be determined. 
$\si'$ and $\om'$ are composed of {\it odd} powers of
tanh($k\rho$), whereas $P$ is of {\it even} powers. 
They come from the behavior at the ultra-violet region
(\ref{sol8b}).
$k$ is a new scale parameter which shows the {\it thickness}
of the pole. In the following, we take $k=1$ for simplicity.
\footnote{
This is because eq.(\ref{sol5}) is invariant under the change:\ 
$\rho\ra k\rho,\ \la\ra (1/k^2)\la,\ \La\ra (1/k^2)\La,\ 
\vz\ra\vz$.
}

The key equation for obtaining the solution of the form (\ref{sol9})
is the following expansion formula 
about $1/\rho$. ( The factor $1/\rho$ appears in (\ref{sol5}) ).  
\begin{eqnarray}
\frac{1}{\rho}=\frac{2}{\tanh\,\rho}
\sum_{n=0}^\infty\frac{s_{2n}}{(2n)!}\{\tanh\,\rho\}^{2n}\ ,\ 
\left.\frac{d^{2n}}{dx^{2n}}\left(\frac{x}{\ln\frac{1+x}{1-x}}\right)\right|_{x=0}
\equiv s_{2n}\ ,\nn
0<\rho<\infty\ ,\ 
s_0=\half\com\q s_2=-\third\com\q s_4=-\frac{12}{5}\com\q \cdots\com
\label{app3}
\end{eqnarray}
All coefficients are finite as shown above. 
We can take the limit $\rho\ra\infty$ above, 
and see 
the infinite series {\it converges} and gives 0\ :\ 
$\sum_{n=0}^\infty\frac{s_{2n}}{(2n)!}=0$. 
We expect the coefficient-series appearing in (\ref{sol9})
have the similar behaviors. 
It is a key, in the present treatment of the infinite series,  
that we take the expansion using 
powers of $tanh\,\rho$ not those of
$\e^{-2\rho}$.  
\footnote{
We can {\it not} reexpress the RHS of 
the first equation of (\ref{app3}) as
$1/\rho=\sum_{n=0}^\infty\frac{s'_{2n}}{(2n)!}\e^{-2n\rho}$ 
with {\it finite} coefficients.
The same notice is said about (\ref{sol9}).  
}

We can show the above form of solutions indeed satisfy (\ref{sol5}) 
if the coefficients satisfy some recursion relation\cite{SI0012}.
The first few orders are given as
\begin{eqnarray}
\left\{
\begin{array}{c}
c_1=\frac{1}{4\al}\{ \fourth\la\vz^4(1-\ez^2)^2+\La \}\com\\
d_1=-\frac{2}{3}c_1\com\\
e_0\ :\ \mbox{free parameter},
\end{array}\right.\nn
\left\{
\begin{array}{c}
c_3=\frac{3}{32}\frac{\la^2\vz^6}{\al}\ez^2(1-\ez^2)^2
+c_1(2+5\al c_1)\ ,\\
d_3=-\frac{4}{3}c_1(1+5\al c_1)\com \\
e_2=-\fourth\la\vz^2 \ez(1-\ez^2)\pr
\end{array}\right.\nn
\label{sol12}
\end{eqnarray}
We note one {\it free} parameter $\ez$ appears
\footnote{
This freedom corresponds to, when $\la=0$, the constant translation
symmetry ($P\ra P+\mbox{const.}$) in eq.(\ref{sol5}) .
}
and
 all coefficients are expressed by 4 parameters
($\la,\vz,\La,\ez$). The parameters, however, have
3 constraints.
\begin{eqnarray}
1=\sum_{n=0}^\infty\frac{c_{2n+1}}{(2n+1)!}\com\q
1=\sum_{n=0}^\infty\frac{d_{2n+1}}{(2n+1)!}\com\q
1=\sum_{n=0}^\infty\frac{e_{2n}}{(2n)!}\com
\label{sol11}
\end{eqnarray}
which comes from the boundary conditions (\ref{sol6}) and (\ref{sol7}).
Therefore the present solution is a 1(=4-3) parameter 
family of solutions.

\section{Evaluation of Coefficients and Analytic Result }
We present here the results of concrete values
of c's, d's and e's for an {\it input} value $\ez(=P(0)/\vz)=-0.8$.
We solve constraints (\ref{sol11}) by taking 
the first 7 terms (up to n=6th order). 
The most crucial point of the present solutions 
is to confirm the convergence of the infinite
series 
$\sum_{n=0}^\infty\frac{c_{2n+1}}{(2n+1)!},\ 
\sum_{n=0}^\infty\frac{d_{2n+1}}{(2n+1)!}$, and 
$\sum_{n=0}^\infty\frac{e_{2n}}{(2n)!}$, 
which guarantees the present 
boundary condition. The 6th-order approximation calculation
gives the vacuum parameters as
\begin{eqnarray}
\vz=0.9625\com\q \La=-2.725\com\q \la=18.375\pr
\label{ana1}
\end{eqnarray}
\footnote{
These values should not be confused with numerical
results like Sec.6. They should, in principle, be definitely
determined just like the energy eigenvalues of the hydrogen
atom are fixed by the boundary condition.
}
\footnote{
The constraint criterion is
$
(1-\sum_{n=0}^6\frac{c_{2n+1}}{(2n+1)!})^2+
(1-\sum_{n=0}^6\frac{d_{2n+1}}{(2n+1)!})^2+
(1-\sum_{n=0}^6\frac{e_{2n}}{(2n)!})^2\ <\ 10^{-4}
$.
}
The obtained values of the coefficients are shown in Fig.1.
\begin{figure}
\epsfysize=6cm\epsfbox{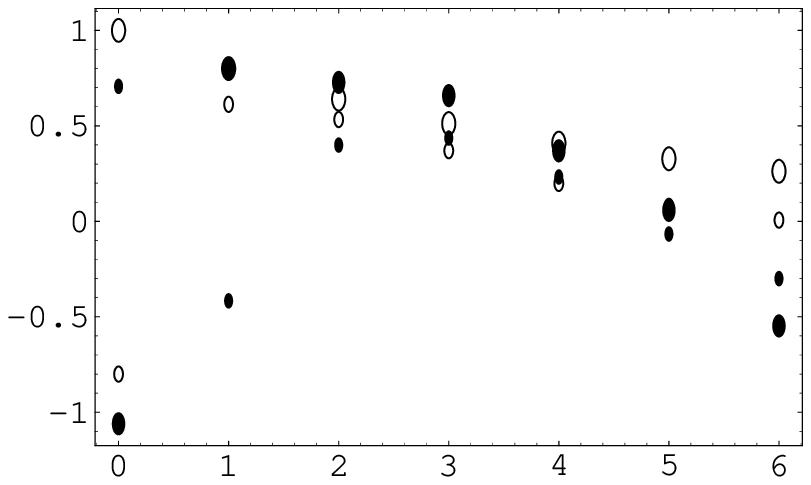}   
   \begin{center}
Fig.1\ 
The values of $c_{2n+1}/(2n+1)!$(large blob),$d_{2n+1}/(2n+1)!$(small blob)
,$e_{2n}/(2n)!$(small circle).
The large circles show $(0.8)^n$. ( $ n=0,1,\cdots,6$ .) 
   \end{center}
\end{figure}
In the figure we also plot the data from
the geometrical series: 
$1/(1-x)=1+x+\cdots$ at $x=0.8$.  
Comparing them,
we can recognize the convergence
of the coefficient-series (up to this approximation order). 
Note that all three series are 'oscillating',
which is advantageous for their convergence. 
Using these results, the analytical results of
$P(\rho),\si'(\rho)$ and $\om'(\rho)$ ( (\ref{sol9}))are shown in Fig.2.
\begin{figure}
\epsfysize=6cm\epsfbox{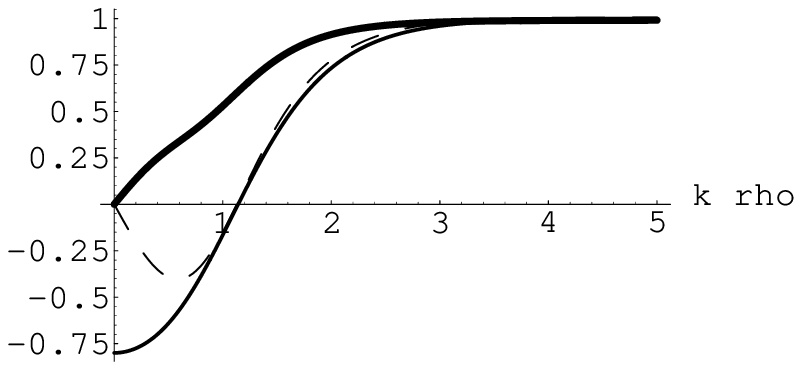}   
   \begin{center}
Fig.2\
The analytic result of $\om'/\al$(bold line), $\si'/\al$(dashed line) and
$P/\vz$(normal line). 
The graphs are depicted by using (\ref{sol9}) with the 6-th order approximation.
The horizontal axis is $k\rho$. We take $k=1$. 
   \end{center}
\end{figure}
We can see comparative behaviors of two 'warp' factors ($\si',\om'$)
in the ultra violet region, $\rho \leq 2$. 
In particular we notice a very 'delicate' dip
in the behavior of the second 'warp' factor $\om'$
at $k\rho\sim 1$. 
The (6D) Riemann scalar curvature is also shown in Fig.3. 
\begin{figure}
\epsfysize=6cm\epsfbox{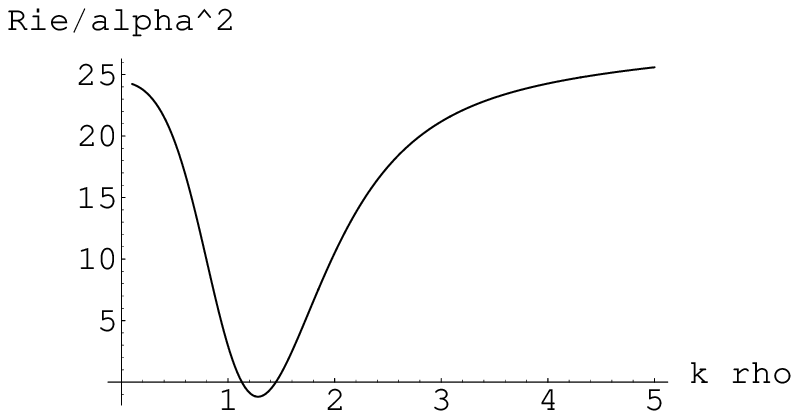}   
   \begin{center}
Fig.3\ 
(6D) Riemann scalar curvature $\Rhat/\al^2$
in the 6th order approximation. 
The horizontal axis is $k\rho$. We take $k=1$.
   \end{center}
\end{figure}
It shows the curvature is positive both inside and outside
of the pole, whereas negative (small absolute value) or zero between them. It
is {\it non-singular everywhere}. There is no horizon in $0\leq \rho<\infty$.
The deficit angle at $\rho=\infty$ is $2\pi$.

\section{Numerical Results}
As the coupled differential equations for 
$P(\rho), \si'(\rho)$ and $\om'(\rho)$, 
the equations (\ref{sol5}) with $m=0$, 
have the standard form
of the numerical analysis, that is, Runge-Kutta method. 
We can numerically solve (\ref{sol5}) {\it without any ansatz}
about the form of the solution.
In this approach, the choice of the initial values with
{\it high precision} is required. In the present case
we cannot take $\rho=0$ as the initial point because
the appearance of the factor $1/\rho$ in (\ref{sol5}).
(Note this does not say the solution is singular at $\rho=0$. 
We seek a {\it non-singular} solution.)
We take $\rho=0.1$ as the initial point of $\rho$. 
As for the initial values we borrow the results from
Sec.5\ :\ 
$P(0.1)=-0.79386,\ \si'(0.1)=-0.10488,\ \om'(0.1)=0.070046$. 
The result by the Runge-Kutta calculation is shown in Fig.4. 
\begin{figure}
\epsfysize=4cm\epsfbox{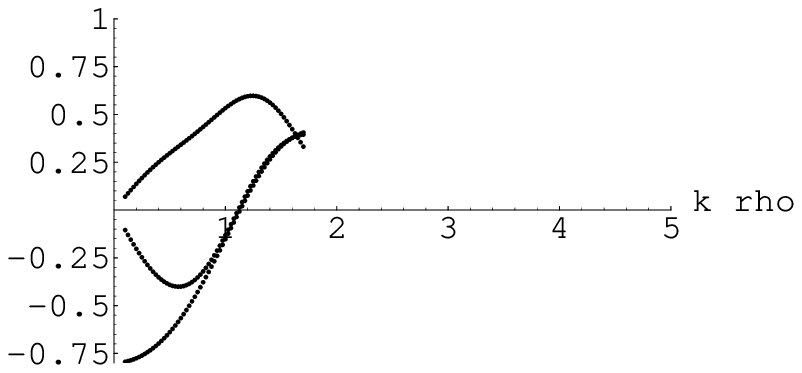}
   \begin{center}
Fig.4\
The numerical results for $\om'/\al$(top), $\si'/\al$(middle) and
$P/\vz$(down). They are obtained by Runge-Kutta method.
One step value along $k\rho$-axis is 0.025. About 65 points are
plotted for each line in the figure. 
The initial point is $\rho=0.1$. 
The horizontal axis is $k\rho$. We take $k=1$.
   \end{center}
\end{figure}
It shows the analytic solution of Sec.5 is well reproduced
in the ultraviolet region ( small-$\rho$ region )
but not for the infrared region. 
We should notice that the the consistent region (with the analytic result) 
, $\rho\leq 1.3$,  
becomes definitely larger than that of the 2nd order approximation
\cite{SI0012} 
, $\rho\leq 0.9$. 
The numerical output data stop at $\rho\sim 1.7$ 
with producing imaginary values.
This occurs because keeping the {\it positivity}, 
${P'(\rho)}^2\geq 0$, 
becomes so stringent
in the infrared region. The quantity becomes
so small in that region and vanishes at $\rho=\infty$. 
We understand that further higher precision is required
for the initial values 
in order to extend the consistent region furthermore. 
\section{Conclusion}
We have presented the both analytical and numerical
solution in the 6D reduction model. The approximation
oder is $n=6$. Compared with the previous result
of $n=2$ order one, the results improve definitely.
Much evidence about the convergence of the coefficients
series is obtained.  An exact solution, 
for the no potential case, is also obtained.

We notice rather stable behavior 
of the following values
under the increase of the approximation order\ :\ 
the vacuum parameters ($\vz,\La,\la$);\ 
the coefficients ($c_{2n+1},d_{2n+1},e_{2n}|n=0,1,\cdots,6$);\ 
the form of $\si',\om'$ and $P$. This strongly
indicates the result of Fig.2 is quite near the exact result.

\vs 1
\begin{flushleft}
{\bf Acknowledgment}
\end{flushleft}
The author thanks P. van Nieuwenhuizen, M. Rocek and
R. Shrock for discussions or comments
about some points related to this work.
Comments by A. Goldhaber, S.T. Hong, 
Z. Kakushadze, I. Oda, S. Vandoren
and V. Zhukov are appreciated. 
He also thanks R. Roiban and C.M. Hung
for helping him in relation to this work.  
The financial support by the governor of
Shizuoka prefecture is greatly acknowledged. 
Finally he expresses gratitude to 
the hospitality at the C.N. Yang Institute for Physics, 
State University of New York at Stony Brook where this work has been done.
\vs 1


\end{document}